\documentclass[fleqn,usenatbib]{mnras}

\usepackage[T1]{fontenc}
\usepackage{ae,aecompl}

\usepackage{graphicx}	
\usepackage{amsmath}	
\usepackage{amssymb}	

\usepackage{newtxtext,newtxmath}
\usepackage{mathptmx}

\usepackage{pdflscape}
\usepackage{longtable}
\usepackage{supertabular}

\usepackage{rotating}
\usepackage{footnote}
\usepackage{times}

\usepackage{tikz,xcolor,hyperref}

\definecolor{lime}{HTML}{A6CE39}
\DeclareRobustCommand{\orcidicon}{
	\begin{tikzpicture}
	\draw[lime, fill=lime] (0,0) 
	circle [radius=0.16] 
	node[white] {{\fontfamily{qag}\selectfont \tiny ID}};
	\draw[white, fill=white] (-0.0625,0.095) 
	circle [radius=0.007];
	\end{tikzpicture}
	\hspace{-2mm}
}

\foreach \x in {A, ..., Z}{\expandafter\xdef\csname orcid\x\endcsname{\noexpand\href{https://orcid.org/\csname orcidauthor\x\endcsname}
			{\noexpand\orcidicon}}
}

\newcommand{\ovi}{O\,{\sc vi}}

\newcommand{\ha}{H$\alpha$}

\newcommand{\neoniii}{[Ne\,{\sc iii}]}

\newcommand{\helium}{He\,{\sc i}}
\newcommand{\heliumb}{He\,{\sc ii}}

\newcommand{\oxygeniii}{[O\,{\sc iii}]}

\newcommand{\ironvii}{[Fe\,{\sc vii}]}

\newcommand{\ironiii}{[Fe\,{\sc iii}]}

\newcommand{\carboniv}{C\,{\sc iv}}

\def\vhel{\ifmmode{V_{{\rm HEL}}}\else{$V_{{\rm HEL}}$}\fi}
\def\vsys{\ifmmode{V_{\rm sys}}\else{$V_{\rm sys}$}\fi}
\def\kms{\ifmmode{~{\rm km\,s}^{-1}}\else{~km~s$^{-1}$}\fi}
\def\vlsr{\ifmmode{v_{\rm lsr}}\else{$v_{\rm lsr}$}\fi}

\title[Where are the missing symbiotic stars?]{Where are the missing symbiotic stars? Uncovering hidden Symbiotic Stars in public catalogues}

\author[Stavros Akras]{
Stavros Akras $^{1\orcidA{}}$~\thanks{E-mail: stavrosakras@gmail.com, stavrosakras@noa.gr}
\\
$^{1}$Institute for Astronomy, Astrophysics, Space Applications and Remote Sensing, National Observatory of Athens, GR 15236 Penteli, Greece\\
}

\date{Accepted XXX. Received YYY; in original form ZZZ}

\pubyear{2022}

\begin{document}
\label{firstpage}
\pagerange{\pageref{firstpage}--\pageref{lastpage}}
\maketitle

\begin{abstract}
Theoretical predictions of the population of Galactic symbiotic stars (SySts) are highly inconsistent with the current known population. Despite intense effort over the past decades, observations are still far below the predictions. The majority of known SySts so far are identified based on selection criteria established in the optical regime. The recent discovery of SU Lyn with very faint optical emission lines uncloaked a subgroup of SySts with accreting-only white dwarfs. In this particular case, the luminous red giant may overshadow the dimmed white dwarf companion. A new approach to search for this subgroup of SySts is presented, employing {\it GALEX} UV and {\it 2MASS/AllWISE} IR photometry. The FUV-NUV colour index is an  indicator, direct or indirect, for the presence of hot compact companions. The cross-match of the Catalogue of Variable Stars~III obtained from the All-Sky Automated Survey for SuperNovae with the {\it GALEX, 2MASS} and {\it AllWISE} catalogues result in a sample of 814 potential SySt candidates. From them, 105 sources have photometric measurements from both FUV and NUV bands and 35 exhibit FUV-NUV$<$1, similar to what it is expected from known SySts. Five known SySts are recovered, while two new genuine SySts are discovered in spectroscopic follow-up observations after the detection of the typical emission lines. 
\end{abstract}




\section{Introduction}
Theoretical predictions of the population of symbiotic stars (SySts) in the Milky Way are strongly contradicted by the observations of an order of 2–3 magnitudes due to the assumptions considered for the evolution of binary systems, for which very little was known 20–30 years ago. 

The first estimation of the SySts's population in the Milky Way was 4$\times$10$^3$ \citep{Kenyon1986}. In 1992, the total number of SySts was computed around 3$\times$10$^5$ by \cite{Munari1992}, and a year later \cite{Kenyon1993} reduced it to 3.3$\times$10$^4$ SySts\footnote{The number of 3.3$\times$10$^4$ corresponds only to S-type SySts, while the D-type with a Mira companion are $\sim$0.6$\times$10$^4$.}. \cite{Magrini2003} recalculated the population of Galactic SySts of the order of 4$\times$10$^5$ assuming that a 0.5~percent of the population of red giants (RGs) and AGB stars should be in orbit around white dwarfs (WDs). More recently, \cite{Lu2006} refined the Galactic population of SySts between 1.2$\times$10$^3$ and 15$\times$10$^3$ employing a population synthesis model. Discrepancies among theoretical studies are associated to the different assumptions on the SySts's lifetime and/or the evolutionary paths of binary systems that eventually become SySts.

From the comparison of these theoretical predictions with the most recent census of known SySts in the Milky Way \citep[][ and references therein]{Akras2019a,Merc2019} and the latest discoveries over the past few years \citep{Merc2020,Akras2021,Merc2021a,Merc2021b,Merc2022,Munari2021,Munari2022,DeKishalay2022,Petit2023}, it is evident that many Galactic SySts are still missing and waiting to be discovered. Thus far, the vast majority of the known SySts if not all of them, have been discovered and identified following a number of criteria based on optical observations: (i) the detection of some typical emission lines (e.g., \ironvii\ $\lambda\lambda$5727,6087), \ha, \oxygeniii$\lambda$5007, and \heliumb\ $\lambda$4686, (ii) the absorption molecular features such a TiO, VO, and CN that ensure the presence of a red giant, and/or (iii) the detection of the \ovi\ Raman-scattered $\lambda\lambda$6830,7088 lines \citep{Mikolajewska1997,bel2000,Akras2019a}.

\begin{figure*}
\vbox{
\includegraphics[scale=0.40]{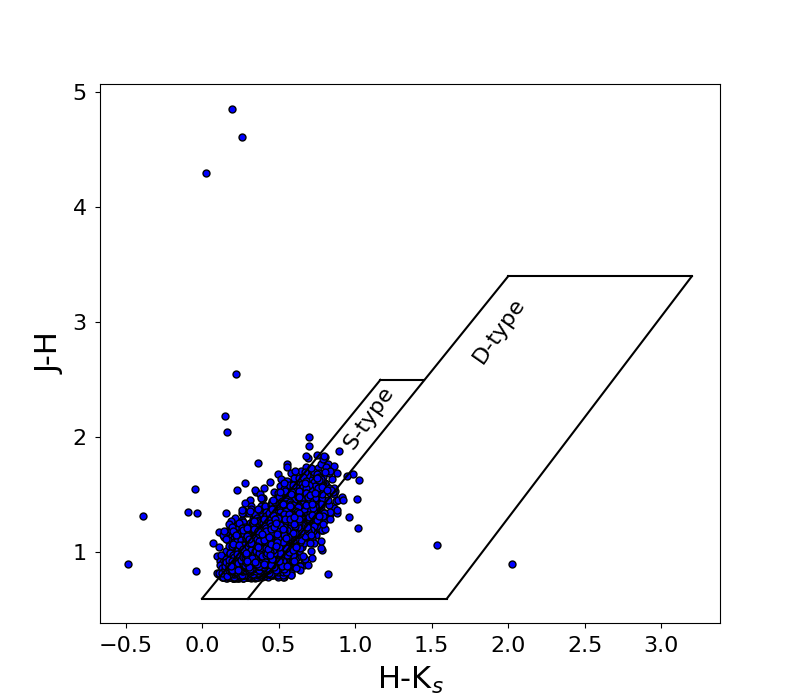}
\includegraphics[scale=0.40]{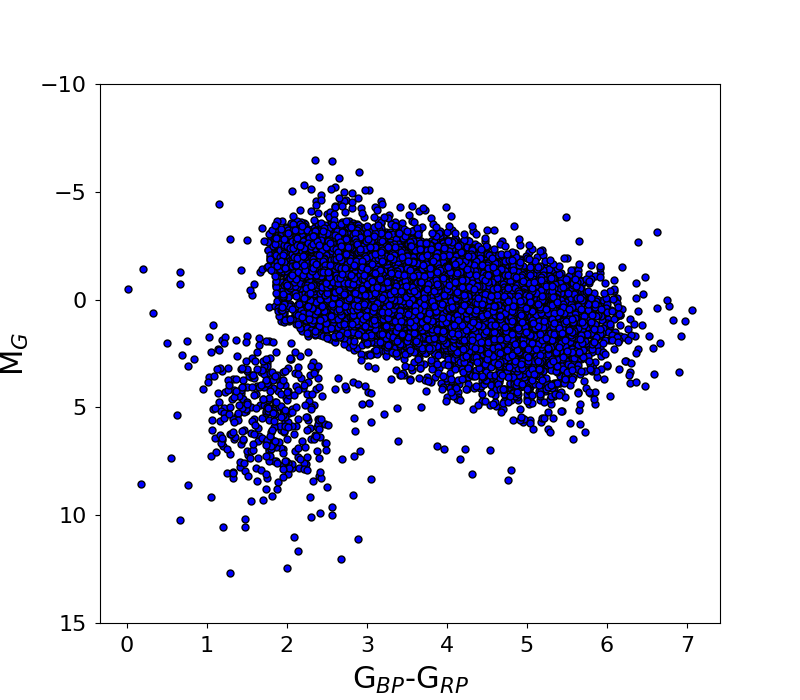}}
\caption[]{Left panel: 2MASS {\it J-H} versus {\it H-Ks} diagnostic diagram of all the sources (29,237) in the ASAS-SN/2MASS/AllWISE cross-matched list. The locus of S and D-type SySts from \citet{Corradi2008,Flores2014,Akras2019b} are also plotted. All sources are concentrated in the regimes of S-type SySts. Right panel: Gaia M$_{\rm G}$ versus G$_{\rm BP}$-G$_{\rm RP}$ HR diagram for the same ASAS-SN/2MASS/AllWISE list. The vast majority has G$_{\rm BP}$-G$_{\rm RP}>$2 and M$_{\rm G}>$5 suggesting evolved stars in the red-giant branch. The smaller group with G$_{\rm BP}$-G$_{\rm RP}<$2.5 and M$_{\rm G}<$2 likely presents some M-dwarfs and main-sequence stars.}
\label{fig1}
\end{figure*}

Although, \cite{Mukai2016} argued for the discovery of the first SySt (SU Lyn) with very weak \ha\ line, whilst other typical emission lines such as \heliumb, \ironvii, \oxygeniii~were absent. The optical spectrum of SU Lyn resembled that of a typical M-type giant star with a weak \ha~emission line, but its hard X-ray emission and UV variability could not be explained without the presence of an accreting hot companion. Note that SU~Lyn satisfies the newly infrared selection criteria of S-type SySts proposed by \cite{Akras2019b,Akras2021}, providing an additional verification of its symbiotic nature. Follow-up X-ray, UV, optical and near-IR observations of SU~Lyn have unquestionably confirmed the symbiotic nature of this accreting-only SySt with M=0.7-0.85~M$\odot$ and T$_{\rm eff}$=37000$\pm$15000~K \citep{Oliveira2018,Vipin2021}. According to the X-ray spectral classification scheme \cite[the $\alpha$, $\beta$, $\gamma$, and $\delta$ scheme, ][ and references therein]{Luna2013}, SU Lyn is classified as  $\delta$-type symbiotic star. An interesting relation between the X-ray spectral types of SySts and the \ovi~$\lambda$6830~Raman-scattered line has been reported by \cite{Akras2019a}. In particular, \ovi~$\lambda$6830 line is only detected in $\alpha$ and $\beta$-type SySts for which the X-ray emission is associated with shell-burning WDs resulting in highly ionized gas and strong optical emission lines.

Based on the intriguing discovery of SU~Lyn, \cite{Mukai2016} argued that the current catalogues of SySts are biased to those with more luminous WDs capable of producing strong optical emission lines and a highly ionized circumstellar envelope. {\it Where are the missing SySts with low luminosity WD companions?} It is very likely that many low-luminosity SySts are misidentified and may be listed in catalogues of typical red giants or even in catalogues of variable stars or emission line sources.

A new accretion-only SySt, THA~15-31, has recently been discovered by \cite{Munari2022}. THA~15-31 displays only Balmer lines in its optical spectrum, molecular TiO bands and strong UV-excess, but no X-ray emission. Note that this newly discovered SySt also satisfies the IR-selection criteria of S-type SySts \citep{Akras2019b}.

Besides SySts, the family of WD~+~RG binary systems also includes Barium and Technetium-poor extrinsic S stars or more general chemically-peculiar stars (hereafter CPSs). They do not show symbiotic activity \citep[][and reference therein]{Jorissen2002} and do not satisfy the optical criteria of SySts. Nevertheless, it may be possible that SySts with low luminosity WDs or feeble symbiotic activity are hidden in these classes of binary systems. The parameters responsible for triggering the symbiotic activity or s-process elements enhancement due to the mass transfer in these binary systems are discussed by \cite{Jorissen2002}. Should both CPSs and SySts be included in the general population of WD~+~RG binary systems when they are compared with the theoretical predictions in the Milky Way?

In this paper, I search for potential SySt candidates camouflaged as regular stars without any prior information coming from optical spectroscopy, but combining information from ultraviolet \citep[GALEX,][]{Bianchi2011,Bianchi2017}, near-infrared \citep[2MASS,][]{Cutri2003} and mid-infrared \citep[WISE, ][]{Wright2010,Cutri2014}. Information from the All-Sky Automated Survey for Super-Novae \citep[ASAS-SN,][]{Shappee2014} and Gaia \citep[][]{Gaia2016,Gaia2018} are also combined in order to better constrain the list of SySt candidates. The paper is organized as follows: Section 2 hosts the methodology followed to search for new candidates which display the representative characteristics of the presence of red giants and hot WDs or an accretion disk around it. The results of the GALEX and 2MASS/AllWISE analysis as well as the follow-up spectroscopic discovery of two new genuine SySts are presented in Sections 3 and 4, respectively. The difference in FUV-NUV colour index between SySts and single dwarfs, red giants, white dwarfs, Barium and Technetium-poor stars is discussed in Section 5. I end up with the conclusions in Section 6.

\begin{figure*}
\vbox{
\includegraphics[scale=0.365]{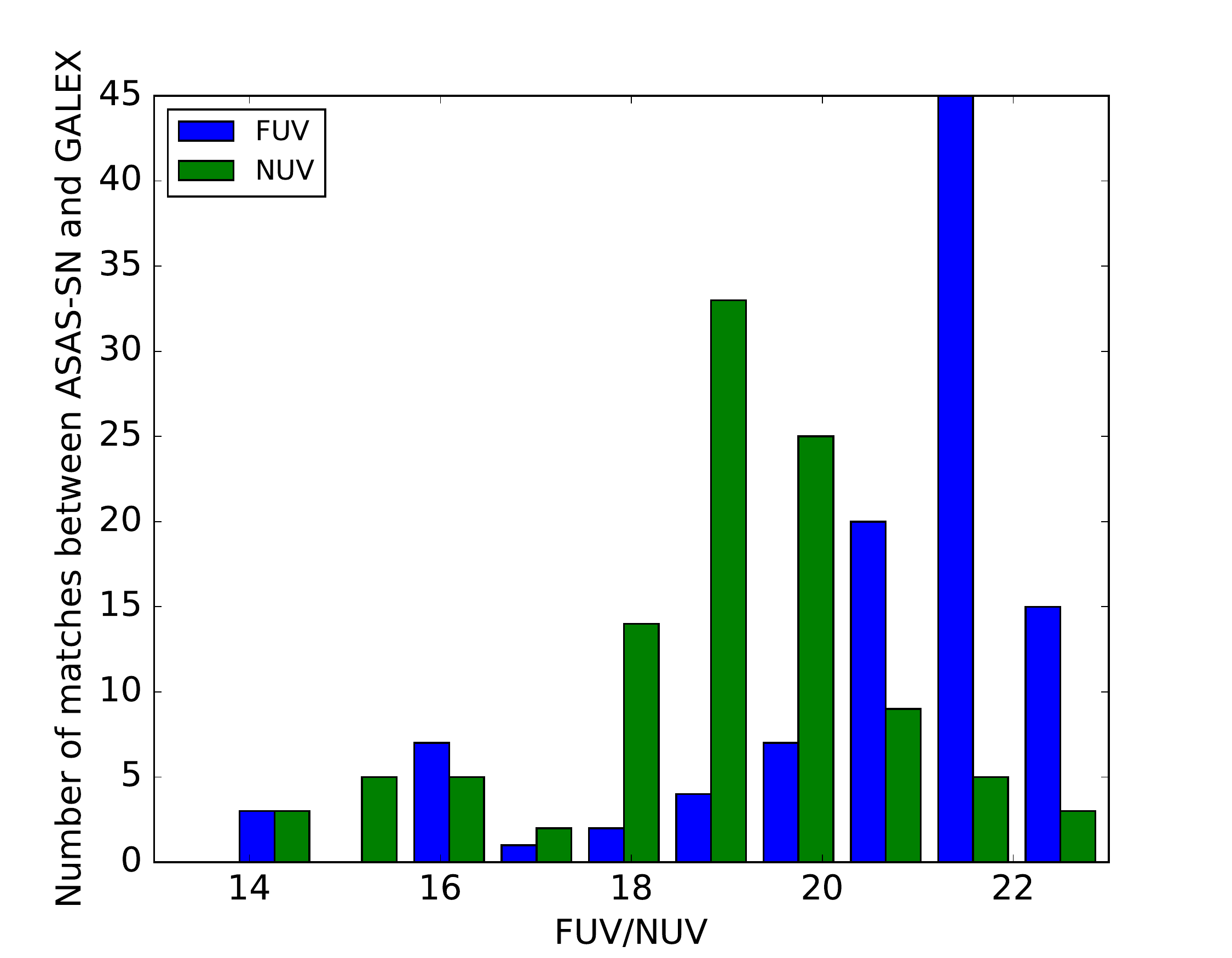}
\includegraphics[scale=0.365]{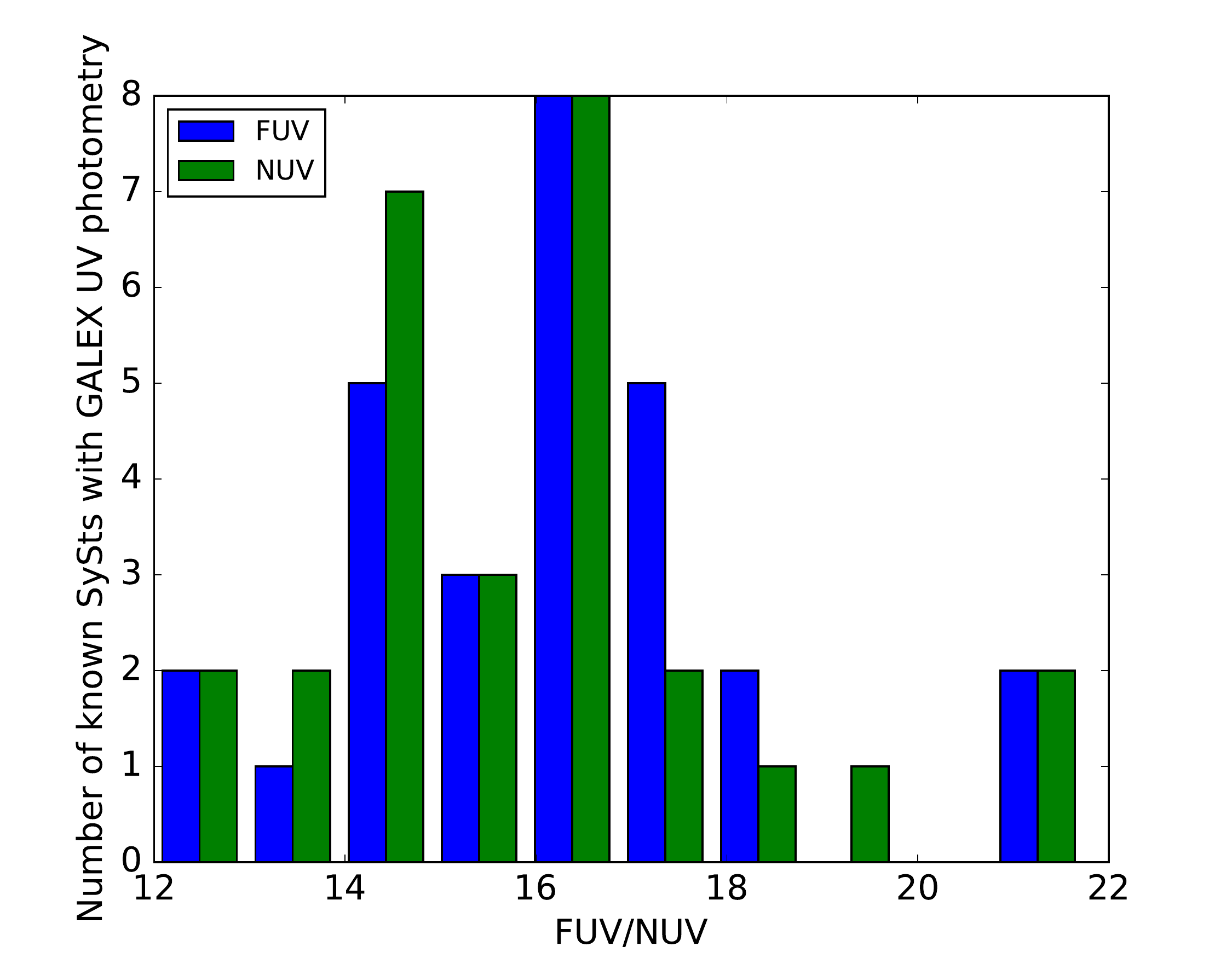}
\includegraphics[scale=0.365]{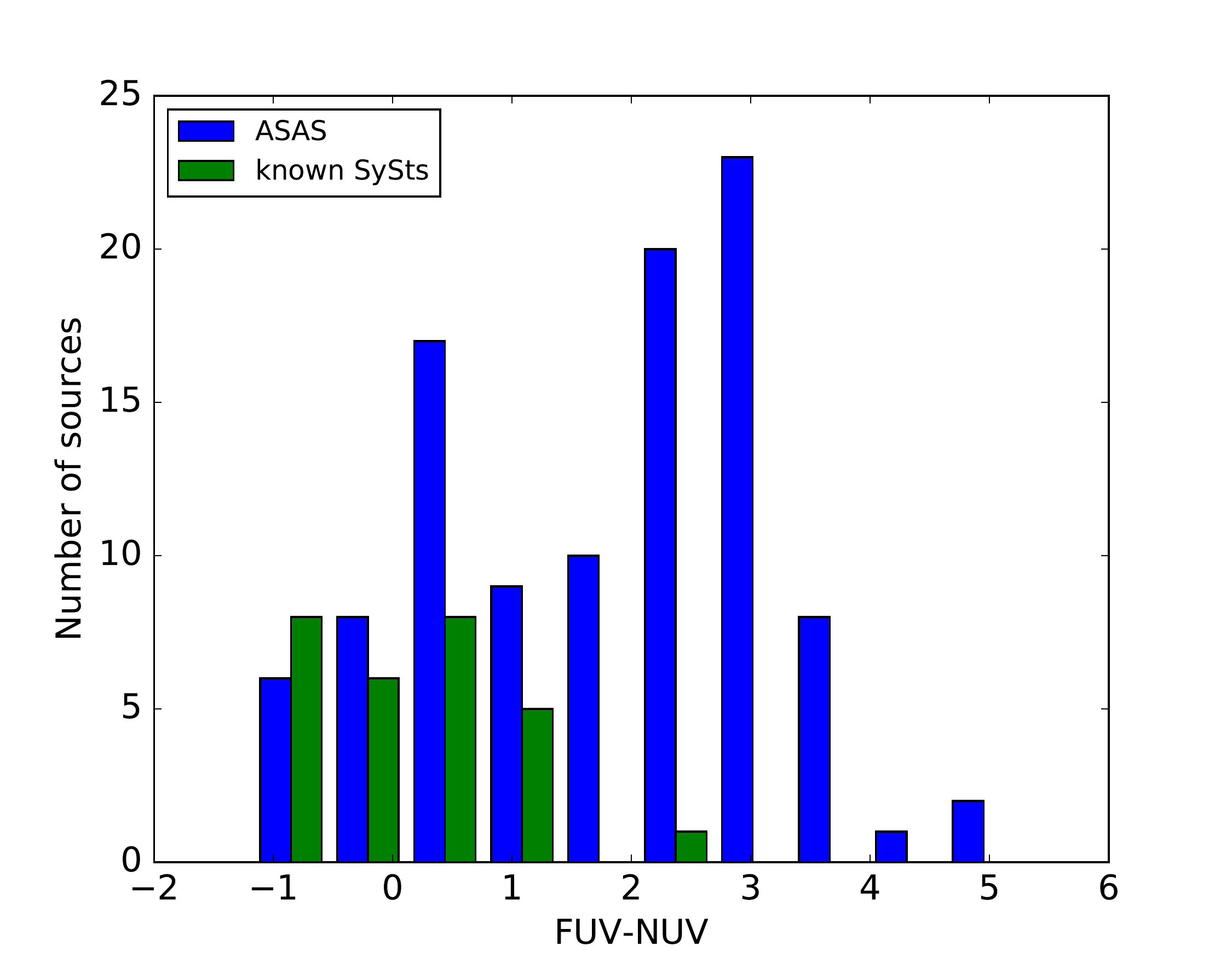}
\includegraphics[scale=0.365]{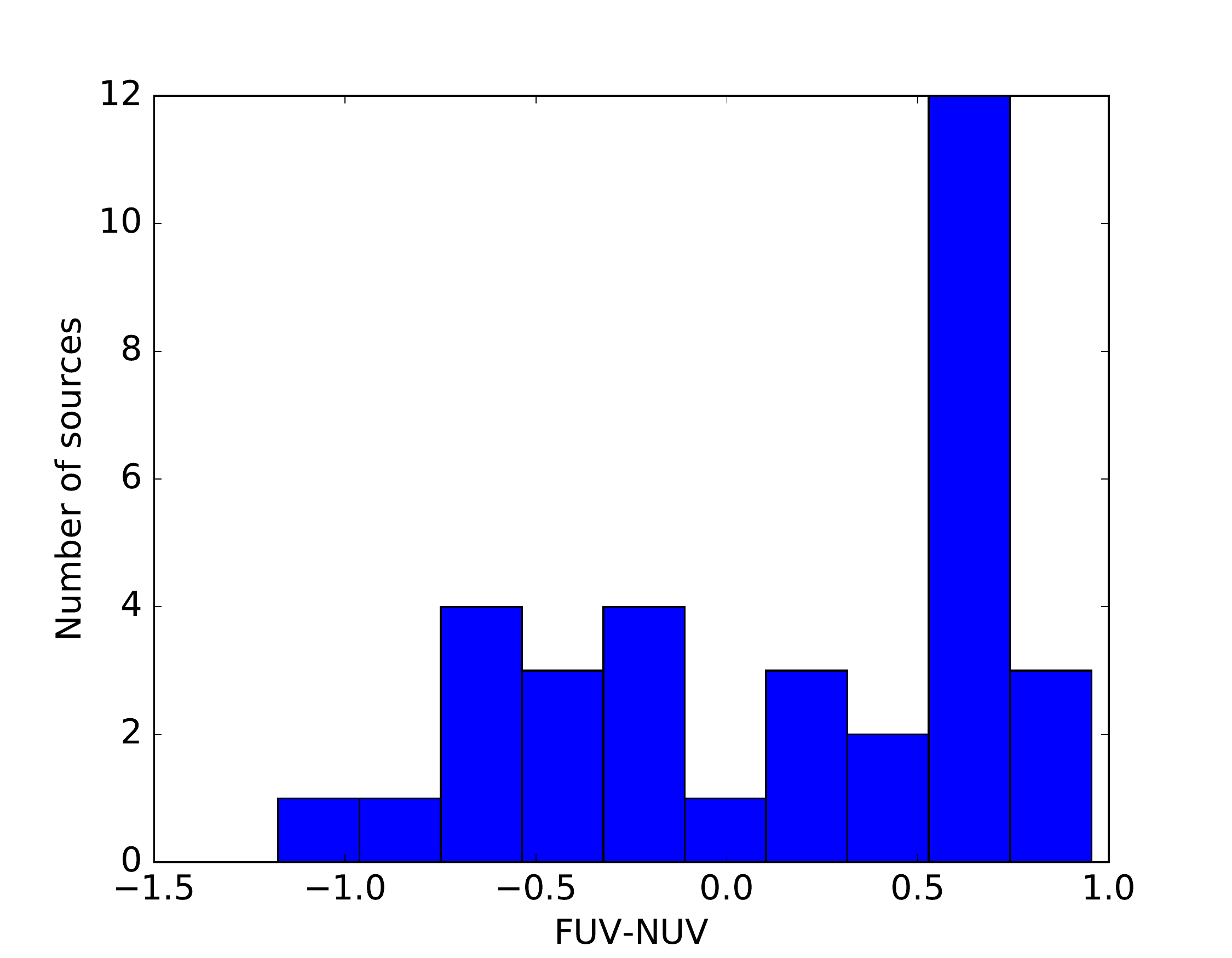}}
\caption[]{The distribution of {\it GALEX} UV photometry of the sources in the {\it GALEX/ASAS-SN} (upper left panel) and the {\it GALEX/Known-SySts} (upper right panel) lists. The lower left panel displays the distribution of the FUV-NUV colour index of the sources in the {\it GALEX/ASAS-SN} and {\it GALEX/Known-SySts} lists. The lower right panel presents the distribution only for those source with FUV-NUV$<1$.}
\label{fig2}
\end{figure*}

\section{Methodology}
The recent discovery of SU~Lyn with weak \ha\ emission, absent \oxygeniii~and \heliumb\ lines, and obvious UV-excess \citep{Mukai2016} in conjunction with the most recent discovery of THA~15-31 \citep{Munari2022} have inspired me to search for those red giants that display strong UV-excess and at the same time satisfy the IR selection criteria of S-type SySts \citep{Akras2019b,Akras2021}. The methodology is divided into two steps. First, it identifies the stars that exhibit characteristics of red giants satisfying the IR selection criteria and then those stars that show strong evidence for the presence of a hot component, either directly due to the photospheric emission from WDs or indirectly from the emission of an accretion disk around WDs. Each step is described in more details in the following subsections.

\subsection{Red giant companion}

In this first step, we make use of the Catalogue of Variable Stars~III\footnote{The catalogue was downloaded in the beginning of 2020. Since then, three more catalogues have been published. The most recent update was made on 08/10/2021, and it lists almost 71,000 more variable objects. \citep{Jayasinghe2018a,Jayasinghe2018b} from the All-Sky Automated Survey for SuperNovae\footnote{https://asas-sn.osu.edu/variables} \citep[ASAS-SN,][]{Shappee2014}. The ASAS-SN survey covers the entire visible sky with telescopes in both hemispheres to a depth of $<$17~mag in the V band and saturation level of $\sim$10-11~mag. Each field of the ASAS-SN survey has 200–600 epochs of observations.}
The catalogue contains $\sim$412,000 variable stars classified as $\delta$ Scuti, RR Lyrae, Cepheids, rotational variables, eclipsing binaries, semiregular and irregular variables and Mira variables. The semiregular and irregular variables classes include red giants, and they are of particular interest. 

The IR (2MASS/AllWISE) selection criteria of S-type SySts \citep{Akras2019b,Akras2021} were applied to the Catalogue of Variable Stars~III resulting in a sample of 96,808 variable sources. To further filter the sample, we used the IR criteria between SySts and K/M-type single giants from \cite{Akras2019b} which reduced the sample to 64,730 stars. According to \cite{Gromadzki2013}, SySts stars display short-term variabilities between 50 and 200 days, besides the long-term and orbital variations, associated with the stellar pulsations of the giant companions. A variability period cut-off was applied to our sample in order to keep only those sources with period $>$50 days. This resulted in our final ASAS-SN/2MASS/AllWISE cross-matched list of 29,237 stars.

In Figure~\ref{fig1}, we present the distribution of the sources in the ASAS-SN/2MASS/AllWISE list in the typical {\it J -- H} versus {\it J -- Ks} diagnostic colour-colour diagram (DCCD, left panel) \citep{Corradi2008,Flores2014,Akras2019b} and in the Gaia M$_{\rm G}$ versus G$_{\rm BP}$-G$_{\rm RP}$ Hertzsprung--Russell diagram (right panel). The vast majority of the sources are well-placed in the locus occupied by the S-type SySts in the near-IR DCCD. Yet, several other classes of stars such as single K/M type giants, Mira stars, T Tauri, young stellar objects (YSO), may exhibit the same colour indices \citep[see fig.~1 in][]{Akras2019b}. Nevertheless, the application of the IR selection criteria is expected to have reduced substantially the contamination from these classes. 

\begin{figure*}
\vbox{
\includegraphics[scale=0.37]{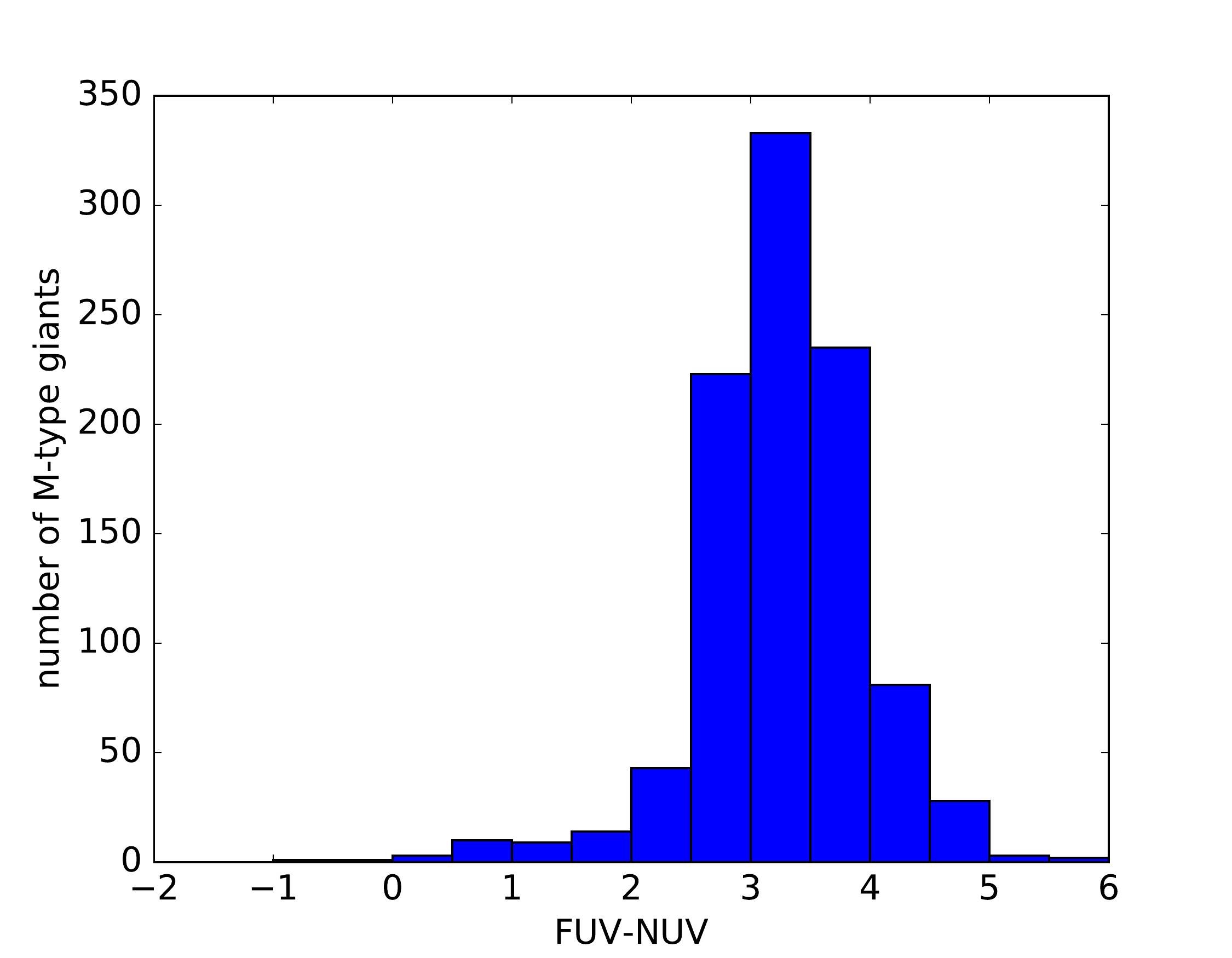}
\includegraphics[scale=0.37]{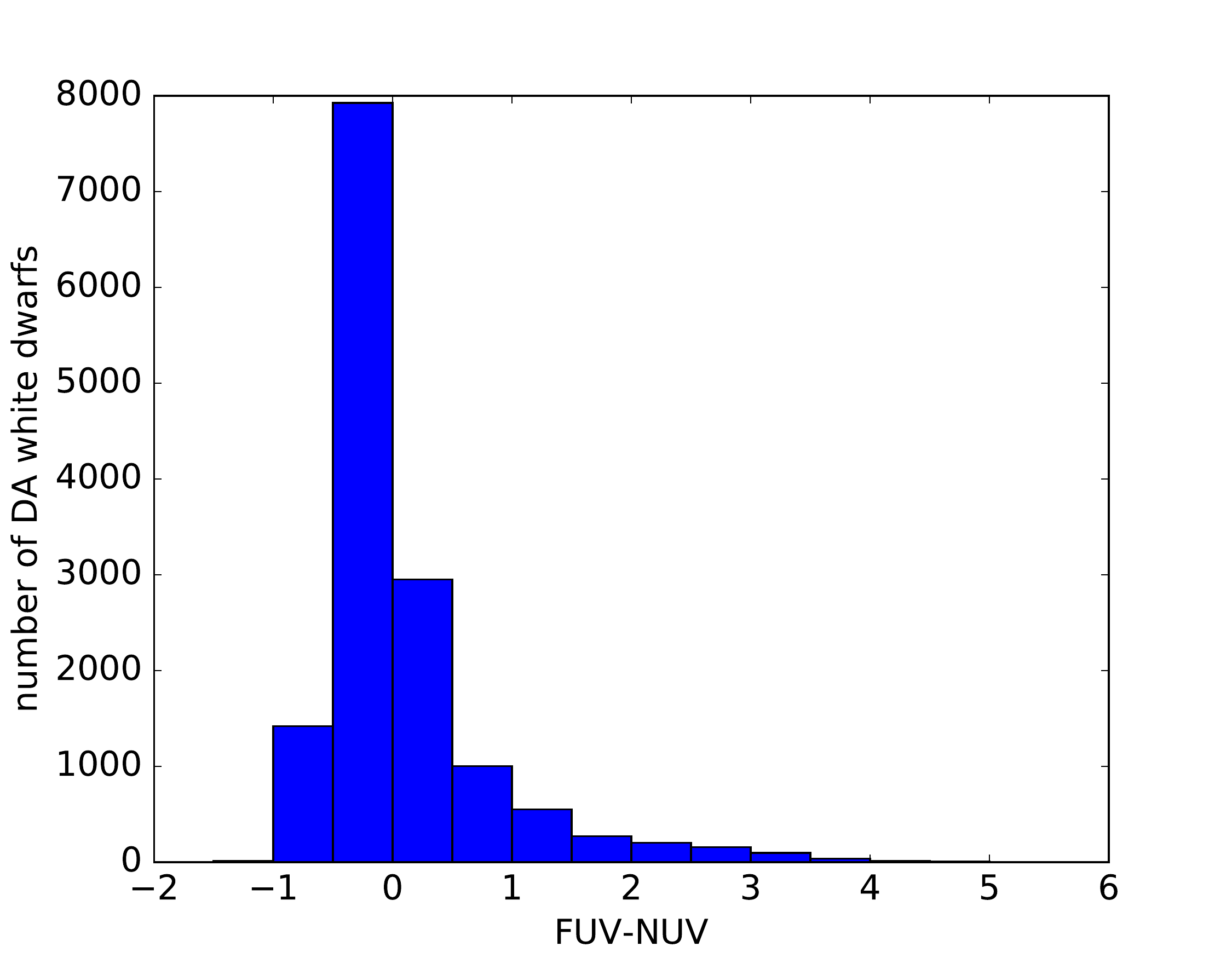}}
\includegraphics[scale=0.375]{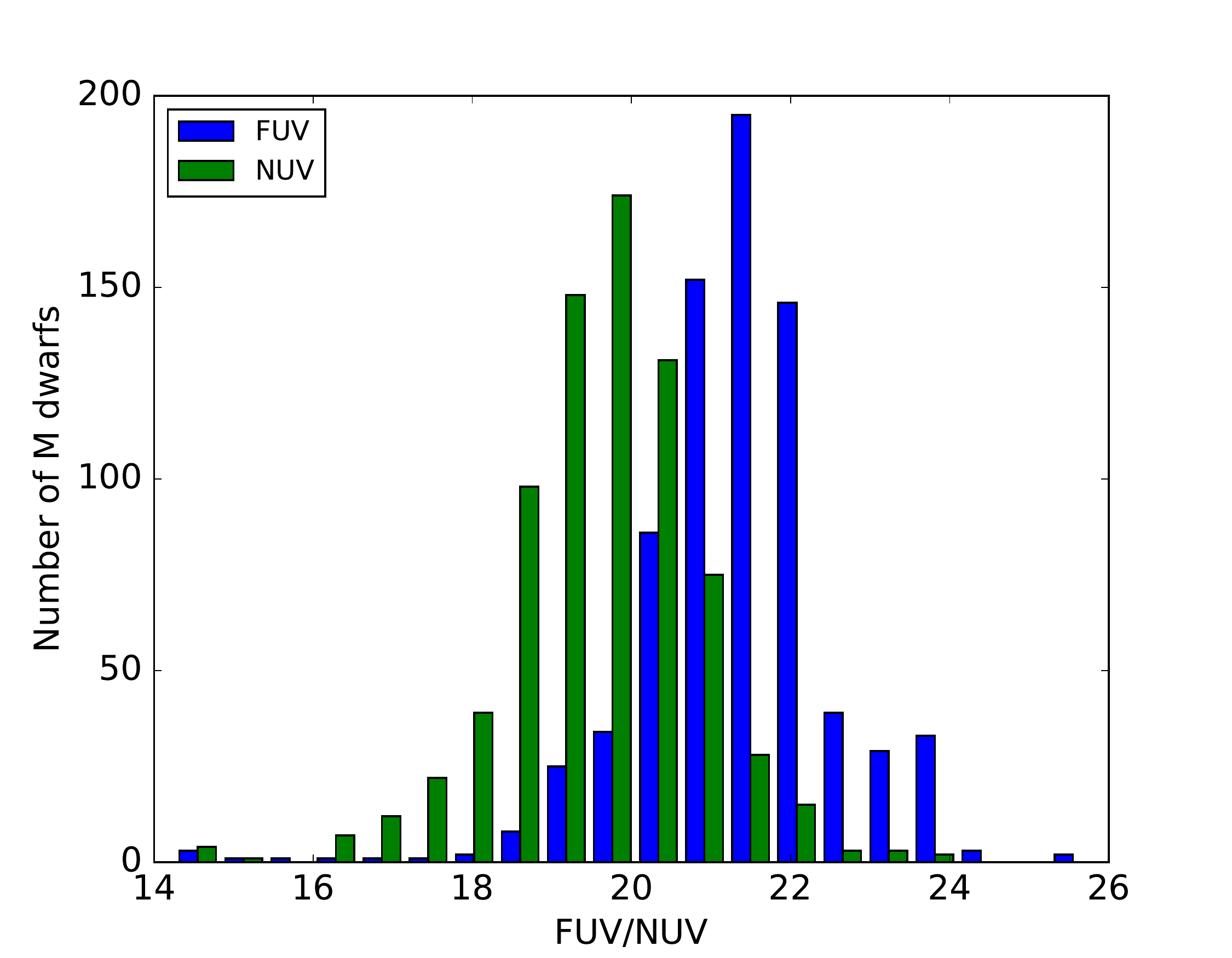}
\includegraphics[scale=0.375]{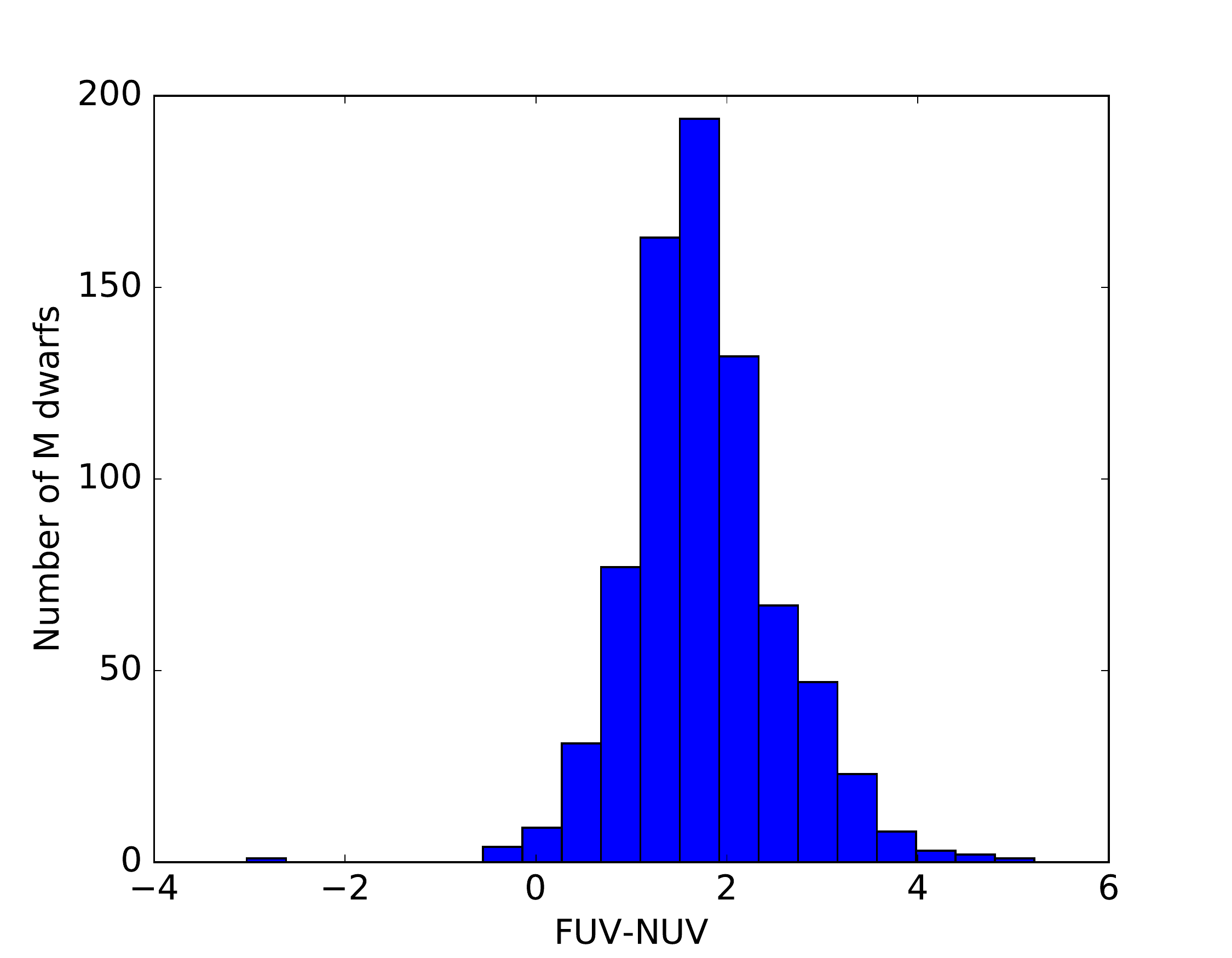}
\caption[]{Upper panels:FUV-NUV colour index distribution of M-type giants (left panel) and DA white dwarfs (right panel). Lower Panel: FUV/NUV photometry (left panel) and FUV-NUV colour index (right panel) distributions of M-type dwarfs}
\label{fig3}
\end{figure*}

It should be noted that M-type dwarfs stars were excluded from the training sample in \cite{Akras2019b} because of their weak \ha\ emission and the violation of the IPHAS criterion of SySts \citep{Corradi2008}. In view of the fact that we do not take into account the IPHAS criterion or any \ha-criterion, variable M dwarfs may be still present in our list.

According to the Gaia M$_{\rm G}$ versus G$_{\rm BR}$-G$_{\rm RP}$ HR diagram, the majority of the sources in the ASAS-SN/2MASS/AllWISE list are evolved stars in the red giant branch phase, G$_{\rm BP}$-G$_{\rm RP}>$2 and M$_{\rm G}>$5, with very few contaminants from T~Tauri, YSO, dwarfs stars or main sequence stars, G$_{\rm BP}$-G$_{\rm RP}<$2.5 and M$_{\rm G}<$2 (see also fig.~22 in \cite{Jayasinghe2018b}).

\subsection{White dwarf companion}  
The analysis in the previous section ensures that the majority of the sources exhibit signatures of red giant stars, and thus they are potential cool companions in binary systems. In the next step, we search for those stars from the previous step that display either direct signatures for the presence of a hot WD companion or indirect signatures from the UV emission of an accretion disk around the WD companion. For this exercise, we make use of the revised catalogue of {\it GALEX} UV sources \citep[GUVcat\_AIS, ][]{Bianchi2017}.

The {\it GALEX} survey mapped an area of $\sim$25000 square degrees with two bands. The far-UV band (FUV) has $\lambda_{\rm eff}\sim$1258\AA\ covering the spectral range from 1340 to 1806\AA, and the near-UV band (NUV) has $\lambda_{\rm eff}\sim$2310\AA\ with a spectral range from 1693 to 3007\AA. {\it GALEX} has successfully been used to find new single WD or in binary systems \citep[e.g.][]{Bianchi2011}. Therefore, the unique photometry of {\it GALEX} will be valuable for the mining of potential WD companions, either directly or indirectly. The GUVcat\_AIS {\it GALEX} catalogue lists 82,992,086 unique UV sources, and it is constructed based on the observations from the All Sky Survey (AIS) with both detectors (FUV and NUV) on. This means that UV sources in the GUVcat\_AIS catalogue without available FUV photometry were below the detection limit \citep{Bianchi2017}. It should also be noted that AIS observations do not cover the Galactic disk, and thus the majority of the Galactic stellar sources are not observed. The results of this work represent only a part of the stellar population in the Milky Way. 

The cross-matching of the ASAS-SN/2MASS/AllWISE list with the GUVcat\_AIS {\it GALEX} catalogue resulted in a list of 814 matches, considering a radius of 5~arcsec due to the FWHM of the {\it GALEX} PSF of 4.2 and 5.3~arcsec for the FUV and NUV filters.

\begin{scriptsize}
\begin{landscape}
\centering
\tablecaption{Table1: List of potential candidate SySts with UV photometry}
\renewcommand{\arraystretch}{1.35} 
\label{table1}
\begin{tabular}{llllllllllllllllll}
\hline
ASASSN-V$^a$ & Other name & RA      & Dec.        & FUV   &  NUV  & M$_{\rm G}$ & G$_{\rm BP}$-G$_{\rm RP}$ & SIMBAD & Refs.  & P$_{\rm ASASSN}^a$ & P$_{\rm VSX}^b$ & Distance$^c$ \\     			
         &      & (J2000.0)   & (J2000.0)   & (mag) & (mag)  & (mag)  & (mag)  &      &        & (days)  & (days) & kpc\\   
\hline 
J103437.09-401940.4 & & 158.654896 & -40.327779 & 19.226$\pm$0.090 & 18.543$\pm$0.045 & -0.024 & 3.577 & & & 215.9039612 & 65.4 & 1.12$^{+0.16}_{-0.12}$ \\
J180148.62+813925.5 & 2MASS J18014909+8139252 & 270.455019 & +81.656976 & 20.350$\pm$0.192 & 19.793$\pm$0.084 & -2.509 & 2.931 & LPV & & 548.0866926 & & 3.47$^{+0.53}_{-0.42}$ \\
J080553.51-065032.6 & IRAS 08034-0641 & 121.473753 & -6.84334 & 21.062$\pm$0.318 & 21.621$\pm$0.336 & 0.589 & 3.838 & star &  & 630.6335752 & & 0.92$^{+0.13}_{-0.11}$ \\
J231839.50+182539.0 & IRAS 23161+1809 & 349.663968 & 18.426775 & 22.008$\pm$0.396 & 21.817$\pm$0.279 & -1.226 & 3.699 & star &  &  270.9739143 & & 1.94$^{+0.55}_{-0.38}$\\
J115151.88-422751.3 & V422 Cen & 177.966346 & -42.464231 & 16.343$\pm$0.036 & 16.477$\pm$0.023 & -1.533 & 4.098 & VSM & & 415.8376321 & 193 & 0.83$^{+0.08}_{-0.07}$\\
J085048.16-124517.9 & IRAS 08484-1234 & 132.700944 & -12.755286 & 21.814$\pm$0.349 & 21.316$\pm$0.218 & -1.309 & 3.515 & star & & 671.0213235 & 953 & 1.67$^{+0.31}_{-0.23}$\\
J203557.17+201127.2 & LT Del & 308.988768 & +20.191100 & 15.937$\pm$0.021 & 16.381$\pm$0.017 & -1.262 & 1.662 & SySt & 1 &  530.3445585 & 476 & 5.81$^{+1.72}_{-1.20}$\\
J194225.42-680735.5 & V399 Pav & 295.606744 & -68.126459 & 13.805$\pm$0.007 & 14.662$\pm$0.006 & -2.954 & 2.838 & SySt & 1 & 590.8957458 & 565 & 3.71$^{+0.99}_{-0.68}$\\
J210006.29-423843.7 & DD Mic & 315.026121 & -42.645440 & 14.038$\pm$0.008 & 14.144$\pm$0.006 & -3.333 & 1.693 & SySt & 1 & 418.2553265 & 399 & 6.53$^{+1.59}_{-1.15}$\\
J155930.27+255511.9 & T CrB & 239.875575 & +25.920248 & 16.199$\pm$0.043 & 15.245$\pm$0.009 & -0.735 & 2.419 & SySt & 1 & 883.577155 & 227.6 & 0.81$^{+0.03}_{-0.03}$\\
J060910.32-833528.6 & TYC 9493-59-1 & 092.294499 & -83.591200 & 19.528$\pm$0.155 & 19.906$\pm$0.136 & -0.677 & 1.982 & star & & 138.39 & 138 & 11.0$^{+1.69}_{-1.39}$\\
J172133.84+015011.7 & SS 305 & 260.391045 & +01.836472 & 15.776$\pm$0.021 & 16.405$\pm$0.014 & -2.063 & 2.642 & ELS & 2 & 459.11 & 459 & 7.74$^{+2.63}_{-1.79}$\\
J165002.92-230517.0 & THA 23-30 & 252.512028 & -23.088131 & 18.379$\pm$0.113 & 18.591$\pm$0.076 & -1.282 & 2.887 & ELS & 3 & 334.37 & 334 & 7.85$^{+3.44}_{-2.17}$\\
J165319.30+330958.2 & KO Her & 253.330200 & +33.166149 & 22.285$\pm$0.386 & 22.951$\pm$0.412 & -1.113 & 3.171 & LPV & & 661.6351211 & 61 & 2.27$^{+0.27}_{-0.22}$\\
J115539.67+123447.8 & LEE 107 & 178.915616 & +12.580103 & 19.702$\pm$0.154 & 19.141$\pm$0.078 & -2.245 & 1.948 & Pe & &  294.0876492 & 57.556521 & 3.23$^{+0.65}_{-0.49}$\\
J184639.07-552112.7  & AN Tel & 281.662882 & -55.352938 & 20.783$\pm$0.305 & 19.863$\pm$0.151 & -2.121 & 1.924 & LPV & &  213.0895841 & & 2.93$^{+0.38}_{-0.30}$\\
J022508.61-132400.6 & BD-14 450 & 036.285797 & -13.400167 & 16.477$\pm$0.042 & 15.873$\pm$0.018 & -2.357 & 2.151 & Pe & & 674.3792712 & 73.8 & 3.03$^{+0.42}_{-0.34}$\\
J105143.65-270433.1 & & 162.931781 & -27.075852 & 21.589$\pm$0.286 & 22.765$\pm$0.439 & -1.344 & 2.233 & & &  673.7733942 & 128.50 & 3.79$^{+0.93}_{-0.68}$\\
J210350.55-641008.5 & UCAC2 3609393 & 315.960620 & -64.168942 & 20.733$\pm$0.194 & 20.159$\pm$0.100 & -0.619 & 3.145 & star & &  272.0216292 & 202.101852 & 2.76$^{+0.79}_{-0.53}$\\
J175754.82+312529.4 & OQ Her & 269.476839 & +31.424950 & 21.172$\pm$0.301 & 20.724$\pm$0.176 & -1.762 & 2.609 & VSM & & 371.1020341 & & 2.52$^{+0.22}_{-0.18}$\\
J204427.66+191440.5 &  & 311.115194 & +19.244580 & 21.561$\pm$0.343 & 20.831$\pm$0.188 & -0.679 & 3.517 & & & 622.4804518 & 194.061707 & 1.94$^{+0.51}_{-0.34}$\\
J083048.08+412223.1 & IRAS 08274+4132 & 127.700378 & +41.373186 & 22.014$\pm$0.488 & 21.409$\pm$0.312 & -1.604 & 2.973 & star & & 442.7656954 & & 2.07$^{+0.35}_{-0.27}$\\
J073506.43+113737.3 & $[$WWV2004$]$ J0735065+113736 & 113.777990 & +11.626260 & 20.901$\pm$0.202 & 20.310$\pm$0.125 & -0.341 & 2.941 & LPV & & 992.69815 & & 1.72$^{+0.47}_{-0.31}$\\
J104440.47+192523.0 & EW Leo & 161.168463 & +19.423048 & 17.977$\pm$0.085 & 18.283$\pm$0.037 & -2.168 & 3.785 & LPV & &  551.4496933 & 90.895804 & 0.95$^{+0.10}_{-0.08}$\\
J165236.05+165046.2  & TYC 1521-203-1 & 253.150389 & +16.847076 & 17.618$\pm$0.049 & 16.989$\pm$0.023 & -2.785 & 2.446 & CVS & &  399.3786493 &	17.147459 & 4.63$^{+1.00}_{-0.73}$\\
J213439.96-022057.0 & TYC 5212-477-1 & 323.666496 & -02.348910 & 21.842$\pm$0.447 & 20.965$\pm$0.187 & -1.382 & 3.445 & star & & 530.4592668 & 66.093857 & 1.38$^{+0.24}_{-0.18}$\\
J065955.32-615041.1 &  & 104.980970 & -61.845086 & 21.583$\pm$0.442 & 22.292$\pm$0.381 & -1.643 & 3.053 & & &  517.2080886 & 52.97 & 2.69$^{+0.38}_{-0.30}$\\
J190955.52-764421.4 & UCAC2 614990 & 287.478400 & -76.738797 & 21.465$\pm$0.294 & 21.302$\pm$0.218 & -2.312 & 3.135 & star & & 540.9608249 & 189.87352 & 6.47$^{+1.73}_{-1.25}$\\
J194456.44-230115.4 & UCAC2 22584087 & 296.234872 & -23.021048 & 20.187$\pm$0.161 & 19.897$\pm$0.089 & -1.577 & 3.121 &  star & & 416.2300275 & 64.05056 & 3.33$^{+1.98}_{-1.00}$\\
J005904.44+455222.0 & V520 And & 014.768515 &  +45.872855 & 18.779$\pm$0.077 & 17.751$\pm$0.032 & -0.329 & 3.652 & LPV   & & 73.9352712 & 70.0 & 1.91$^{+0.44}_{-0.31}$\\
170633.94+200805.9 & V455 Her & 256.642073 &  +20.134955 & 20.604$\pm$0.242 & 19.914$\pm$0.119 & -0.871 & 3.418 & LPV   & & 62.9524961 & 57.201015 & 1.89$^{+0.21}_{-0.23}$\\
J195948.16-825237.0 & Hen 3-1768 & 299.949838 & -82.877005 & 14.589$\pm$0.010 & 14.905$\pm$0.009 & -3.252 & 1.818 & ELS/SySt & 4 & 90.4840311 & 52.67 & 7.07$^{+1.25}_{-0.97}$\\
J200938.65-052808.4$^d$ & C* 2863 & 302.411034 & -05.468843 & 20.643$\pm$0.178 & 21.063$\pm$0.092 & -1.678 & 1.975 & carbon & & 55.8698306 &  & 4.29$^{+1.33}_{-0.87}$\\
J131436.41-312510.3$^e$ & V957 Cen & 198.651872 & -31.419652 & 20.511$\pm$0.304 & 18.749$\pm$0.037 & 0.067 & 3.014 & Candidate SySt  & 1 & 89.5268432 & 54.545456 & 1.14$^{+0.27}_{-0.19}$\\  
J122432.81-261408.1 & KV Hya & 186.136647 & -26.235779 & 16.145$\pm$0.029 & 15.598$\pm$0.015 & -2.600 & 2.053 & carbon & & 90.7332789 & 97.4  & 4.99$^{+1.18}_{-0.89}$\\  
J085541.52-200603.7 & BD-19 2567 & 133.923212 & -20.100973 & 21.311$\pm$0.391 & 20.577$\pm$0.121 & -2.015 & 3.459 & star   & & 93.834604 & 331  & 1.08$^{+0.09}_{-0.07}$\\  
\hline
\end{tabular}
\begin{flushleft}
$^a$ The Catalogue of Variable Stars~III \citep{Jayasinghe2018a,Jayasinghe2018b} from the All-Sky Automated Survey for SuperNovae (ASAS-SN, \cite{Shappee2014})\\
$^b$ AAVSO International Variable Star Index \citep[VSX,][]{Watson2006}.\\
$^c$ Geometric distances from \citep{Bailer2018} based on Gaia DR2 parallaxes. The indices correspond to the upper and lower bounds on the confidence interval of the estimated distance.\\
$^d$ This object has received the variable star designation, V2011 Aql, in the GCVS5.1.\\
$^e$ This candidate SySt has a FUV-NUV=1.7608~mag.\\
LPV: long period variable, VSM: variable star of Mira Cet type, ELS:emission-line star, Pe: Peculiar star, CVS: Cepheid variable Star\\
References:(1) \cite{Akras2019a} and references therein, (2) \cite{Stephenson1977}, (3) \cite{lim1964}, (4) \cite{Lucy2018} \\
\end{flushleft}
\end{landscape}
\end{scriptsize}

\begin{figure*}
\vbox{
\includegraphics[scale=0.395]{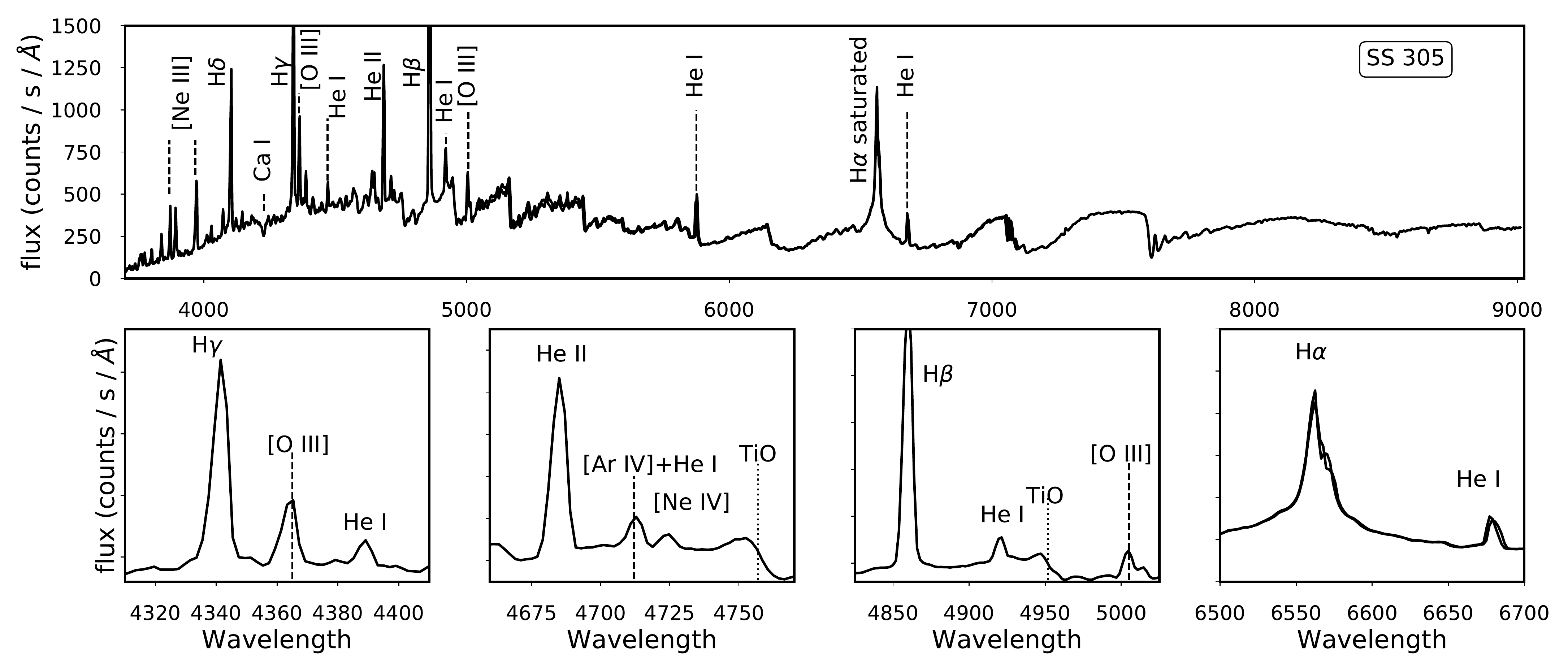}
\includegraphics[scale=0.395]{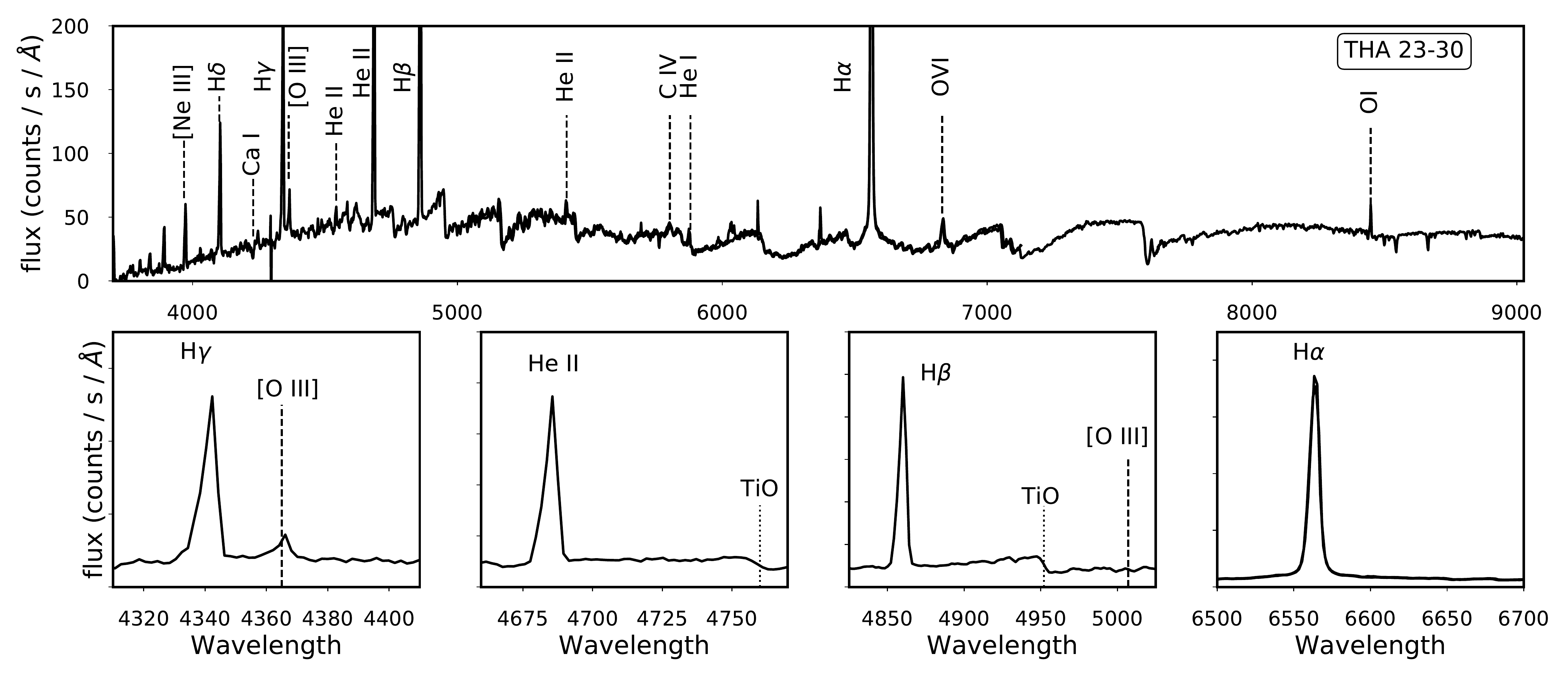}}
\caption[]{Low resolution long-slit spectra of the emission lines sources SS~305 and THA~23-30. The upper panels present the full spectra whilst the lower panels zoom-in the characteristic optical emission lines of SySts. }
\label{fig_spec}
\end{figure*}

The majority of AGNs and QSOs have already been excluded from the ASAS-SN/2MASS/AllWISE list because of the W1-W2 WISE selection criterion. Based on the work by \cite{Akras2019b}, SySts have W1-W2$<$0.1 while QSOs and AGNs have W1-W2 of 1.20 (standard deviation of 0.16) and 0.51 (standard deviation of 0.26), respectively \citep[e.g.][]{Nikutta2014}. QSO are also characterized by FUV-NUV$>$0 (or $>$0.5 for red-shift$>$0.5-0.6) \citep{bianchi2011b, Deharveng2019}. The contamination of our list with background ANGs and QSOs is expected to be small.

From these 814 matches, only 105 have measurements from both UV {\it GALEX} bands, so the FUV-NUV colour index can be explored. Hereafter, we will refer to this list as the {\it GALEX/ASAS-SN} list. Note that the FUV-NUV colour index has not been corrected for interstellar extinction, given that the {\it GALEX} colour is nearly independent for Milky Way dust \citep{bianchi2011b}. As for the remaining 709 candidates without FUV measurements, they are very faint UV sources to be detected in the FUV band, which implies high FUV-NUV colour for these sources.  Moreover, the cross-matching of the list of known Galactic SySts from \cite{Akras2019a} with {\it GALEX} returned 30 matches (hereafter {\it GALEX/Known-SySts} list). 

The distributions of the {\it GALEX} UV photometry as well as the FUV-NUV colour index for both lists, {\it GALEX/ASAS-SN} and {\it GALEX/Known-SySts}, are presented in Figure~\ref{fig2}. The FUV and NUV magnitudes in the {\it GALEX/ASAS-SN} list vary from 14 up to 22~mag with peaks at 21 and 19~mag, respectively (upper left panel). On the other hand, the known SySts ({\it GALEX/Known-SySts} list) are found to exhibit UV photometry between 12 and 21~mag with a peak around 16~mag for both bands (upper right panel). This analysis demonstrates that most of the known SySts, discovered employing the traditional optical criteria, are relatively bright UV sources and certainly brighter than our {\it GALEX/ASAS-SN} candidates. Although, there are some cases of genuine SySts being relatively faint in UV such as StHA~164 (FUV=21.4694$\pm$0.3946, NUV=21.0719$\pm$0.2503) or GH~Gem (FUV=not  measured, NUV=19.6457$\pm$0.1433) implying that UV-faint SySts ($>$19~mag) cannot be ruled out.

A comparison of the FUV-NUN colour index between the {\it GALEX/ASAS-SN} and {\it GALEX/Known-SySts} lists shows that there is a number of {\it GALEX/ASAS-SN} candidates with FUV-NUN colour index comparable to that of genuine SySts (FUV-NUV$<$1, see bottom left panel in Figure~\ref{fig2}). This suggests the presence of a potential hot, compact companion in these candidates. The distribution of FUV-NUV index of the {\it GALEX/ASAS-SN} sources with UV colour index less than 1 is displayed in the bottom right panel of Figure~\ref{fig2}.

Beside the bluer {\it GALEX/ASAS-SN} sources in the FUV-NUV colour, there are also 70 redder sources with 1$<$FUV-NUV$<$5, which may be either SySts like o~Ceti (FUV=15.2035$\pm$0.0200, NUV=12.9255$\pm$0.0044; FUV-NUV=2.278) or even single M-type giant stars with strong chromospheric activity. The upper panels in Figure~\ref{fig3} display the distribution of the FUV-NUV colour for a sample of single red giants and single WDs from the Catalogue of Stellar Spectral Classifications \citep{Skiff2014}. The former show a peak at around 3.5~mag while the latter are clearly bluer with a peak at -0.5~mag. The higher the FUV-NUV colour index, the lower the probability of being a genuine SySts or more general, a WD+RG binary system, but not negligible.

\section{Results: GALEX/ASAN-SN sources}
In Table~\ref{table1}, we list all the 35 {\it GALEX/ASAS-SN} sources with FUV-NUV$<$1, and they are considered as potential candidates of genuine SySts or WD~+~RG binary systems. The first and second columns list the ASAS-SN ID number and other names of the sources. The third and fourth columns give the coordinates (R.A. and Decl.) in the J2000.0 equinox. The {\it GALEX} FUV and NUV photometry as well as the Gaia (G) photometry and the G$_{\rm BP}$-G$_{\rm RP}$ colour index of each source are listed in the fifth, sixth, seventh and eight columns, respectively. Their classification in SIMBAD and references are given in the ninth and tenth columns. Finally, the variability period of each source provided by the ASAS-SN \citep{Shappee2014,Jayasinghe2018a,Jayasinghe2018b} and the International Variable Star Index (VSX) database available at the AAVSO \citep{Watson2006} are also listed in the last two columns. 

Our final {\it GALEX/ASAS-SN} list contains long period variables, variable star of Mira Cet type, emission-line star, peculiar star, Cepheid variable Star, carbon stars, symbiotic stars and some sources without a spectral classification yet. Moreover, significantly different variably periods are provided by ASAS-SN and VSX.

SS~305 and THA 23-30 are of particular interest as they have been classified as emission line stars \citep{Stephenson1977,lim1964}. This means that their symbiotic nature can be verified with optical spectroscopic data.Their periods provided by ASASSN and VSX are in excellent agreement: 459.11/459~days (SS~305) and 334.37/334~days (THA 23-30). In addition to these emission line sources, four known (T CrB, Hen 3-1761, LT Del, CD-43 14304) and one candidate SySt (V957~Cen) are also recovered from the ASAS-SN catalogue. All four genuine SySts are among the brightest UV sources with GALEX photometry $<$16.5~mag and FUV-NUV indices of 0.9545, -0.8567, -0.4443, and -0.1066~mag, respectively.

The SySt candidate, V957~Cen \citep{Akras2019a} is a much fainter UV source with FUV=20.5107~mag and NUV=18.7499~mag. Its FUV-NUV colour index (1.7608) is found to be significantly higher than the known SySts, and therefore it is less likely to be a genuine SySt. 

Hen~3-1768 is also included in the list of emission line stars \citep{Henize1976}. It is a very bright UV source (FUV=14.5894, NUV=14.9054) with FUV-NUV=-0.316, indicative of a potential symbiotic star if we take into account that it satisfies the near-IR selection criteria of S-type SySts \citep{Akras2019b,Akras2021}. The characteristic \ovi\ Raman-scattered line of SySts was detected in this source during the validation phase of the RAMan Search for Extragalactic Symbiotic Stars (RAMSES~II) project \citep[][]{Angeloni2019} and spectroscopically by \cite{Lucy2018} providing the necessary confirmation of its symbiotic nature.

\section{Follow-up spectroscopy of the emission line sources}
The previous classification of SS~305 and THA~23-30 as H$\alpha$-emitters prompted me to obtain their first optical spectroscopic data and unveil their true nature. Low resolution long-slit spectra were obtained on 3$^{rd}$ and 7$^{th}$ of August 2019 using the Goodman High Throughput Spectrograph \citep{Clemens2004} on the 4.1~m telescope at the Southern Astrophysical Research (SOAR) observatory in Chile.

The configuration of the observations covers the wavelength range from 3000 to 9050~\AA\ in two parts: the blue M1 arm from 3000 to 7050~\AA\ and the red M2 arm from 5000 to 9050~\AA\ \citep[see also, ][]{Akras2021}. This wavelength range covers all the characteristic lines (e.g. \ha, \helium, \oxygeniii, \ironiii, \ironvii, \ovi) as well as the TiO and VO molecular bands needed for their classification as genuine SySts. 

The optical spectra of SS~305 and THA~23-30 are shown in Figure~\ref{fig_spec}. Both emission line sources are found to be unquestionable genuine SySts. Hydrogen recombination lines are detected, as expected based on their previous classification as emission line sources \citep{Stephenson1977,lim1964} together with high excitation lines such as \heliumb, \oxygeniii, \ironiii, \neoniii, \carboniv~indicating the presence of a hot companion. The characteristic \ovi\ Raman-scattered line detected mainly in SySts \citep{Allen1980,Akras2019a} is also observed in one of the candidates (THA~23-30). The molecular bands associated with presence of a cold companion are also detected in both sources. Hence, these newly discoveries of SySts demonstrate that combining information/data from the ultraviolet and infrared spectral ranges can also lead to interesting results.

It is worth mentioning that these two H$\alpha$-emitters have significantly lower FUV-NUV colour indices (-0.63 and -0.21, respectively) compared to the colour of SU Lyn (0.32). This can explain the optical emission-line spectra of SS~305 and THA~23-30. This suggests the presence of shell-burning WDs in these newly discovered SySts in contrast with the accreting-only SU Lyn and THA~15-31.

It should also be noted that three more candidates, which are not classified as emission line sources (AN Tel, C* 2863 and UCAC2 614990), were observed without any emission line detected. The FUV-NUV colour index of these three sources is 0.92, -0.42 and 0.16~mag, respectively. The study of their ASAS-SN and {\it GALEX} light curves is required to bring more insights on their classification as accreting-only SySts.

Overall, the proposed methodology provides us with some good SySt candidates for follow-up optical spectroscopic surveys. However, the rate of true positives may not be very high. In this work, five candidates with FUV-NUV$<$1 were observed and only two of them turned out to be genuine SySts (40 per cent). If we restrict the FUV-NUV criterion to $<$0, then 2 out of 3 candidates (66~percent) were turned out to be genuine SySts.

It has to be pointed out that there are several hundreds of common sources between ASAS-SN and GALEX for which FUV measurements are missing. Considering that within the 105 ASAS-SN and GALEX matches with available FUV and NUV photometry, there are 35 that satisfy the FUV-NUV$<$1 criterion (see Table~2), and within these 35 there are seven genuine SySts (five previously known and two newly discovered), we reckon that nearly 50 potential genuine SySts, depending on their FUV-NUV colour index, may exist in the list of 814 ASAS-SN/GALEX matches. 

An additional tool that can also be very useful for the identification of new SySts, and it will be presented in a forthcoming paper, is the investigation of the GALEX light curves of the sources, seeking for evidence of interaction between the two components in potential binary systems. This approach may also reveal potential SySts with low UV WDs and faint or even absent emission lines. 

\section{Discussion}

\subsection{SySts versus single dwarfs, red giants and white dwarfs.}
\cite{Bianchi2007} pointed out that {\it GALEX} sources bluer in the FUV-NUV colour and redder in the optical are likely binary systems with a hot WD and an M-dwarf type star \citep[see also][]{Rebassa2010} or a giant star \citep[see also][]{Li2018}. \cite{Smith2014} carried out a study on the so-called {\it UV-excess sources} which appear to be bluer in the FUV-NUV colour index relative to their spectral type, and possible scenarios, e.g. WDs, subdwarfs, binary systems, and chromospheric activity, for the origin of the UV-excess are discussed by the authors. Only a small part of the {\it GALEX} sources with UV-excess can be explained ($\leq$2~per cent are compact sources; WDs, subdwarfs, $\sim$9~percent are old and young stars with chromospheric activity and $\leq$4~percent are close active binaries).

To ensure that the FUV-NUV colour cut-off results in {\it GALEX} detections for which the UV emission is attributed to the presence of WDs (either from an accretion disc or a stellar wind irradiated by the hot WD photosphere) rather than the chromospheric activity due to magnetic fields on the surface of sinlge giants and dwarfs \citep[e.g.][]{Simon1989,Antova2013,Smith2018}, the {\it GALEX} photometry of a sample of single M-type giants (986 giants) from the Catalogue of Stellar Spectral Classifications \citep{Skiff2014} and a sample of single M-type dwarf (762) selected from the All sky catalogue of bright M dwarfs \citep{lepine2011} were explored. The upper panels in Figure~\ref{fig3} display the FUV-NUV distribution of M-type giants and DA-type WDs and the lower, right panel the distribution of M-type dwarf stars.

The mean FUV-NUV colour index of M-type giants is 3.271~mag with a standard deviation of 0.703~mag, which is consistent with previous studies \citep[see also fig. 1 in ][]{Smith2014}. This analysis also shows that late M-type giants are described by lower FUV-NUV colour index than the earlier types \citep[see fig. 1 in ][]{Smith2014}. As for M-type dwarfs, the mean FUV-NUV colour index is 1.76~mag with a standard deviation of 0.776~mag. A recent study on the chromospheric activity of cold dwarfs has shown that late type dwarfs, with low effective temperature, are characterized by a weaker chromospheric activity \citep{BoroSaikia2018}. Nearly 100 M dwarfs or 13~percent are found to exhibit FUV-NUV$<$1. This number becomes significantly smaller in case a more restricted FUV-NUV$<$0.5 cut-off is applied (23 sources or 3 percent).  

On the other hand, the single DA WDs display a mean FUV-NUV=0.064~mag with a standard deviation of 0.719~mag (14652 WDs) being significantly bluer than both giants and dwarfs. From the sample of DA WDs, we find that 91~percent exhibit FUV-NUV$<$1, while the same percentage for M-giants is only 1.5~percent. This does not necessarily mean that there are no WDs in binary systems with FUV-NUV$>$1. Overall, the identification of binary systems with a WD companion is not feasible by employing only the FUV-NUV colour, as it can also be seen in the figures 5 and 6 in \cite{Bianchi2020}. The FUV-NUV$<$1 colour cut-off is valid for excluding the vast majority of single M-type giant and dwarf stars, but additional data are needed. We also deduce that SySts with weak UV emission ($>$19~mag) are likely misidentified. Based on the sample of known SySts with available {\it GALEX} measurements, genuine SySts can exhibit UV photometry down to 22~mag (right panel in Figure~\ref{fig2}). Recently, a new classical SySt (Hen~3-860), previously classified simply as H$\alpha$ emitter, was discovered in the ASAS-SN catalogue \citep{Merc2022}. This new SySt is not recovered in this work, simply because there is no available GALEX photometric data.

It should also be noted that there are two sources (BY CVn and V934 Her) with FUV-NUV$<$1 in the list of M-type giants used for this analysis, and I searched for further information in public catalogues. V934 Her is a known SySt \citep{Akras2019a,Merc2019} while BY~CVn is classified as MIII-type long-period variable star in various catalogues (All-sky Compiled Catalogue of 2.5 million stars \citep{Kharchenko2001},~JMMC Stellar Diameters Catalogue (JSDC)~Version~2 \citep{Bourges2014}, the Tycho-2 Spectral Type Catalogue \citep{Wright2003}, 2$^{nd}$ Catalogues of Radial Velocities with Astrometric Data \citep{Kharchenko2007}, Extended Hipparcos Compilation (XHIP) \citep{Anderson2012}, All-sky spectrally matched Tycho2 stars \citep{Pickles2010}, Catalogue of Stellar Spectral Classifications \citep{Skiff2014}, General Catalogue of Variable Stars \citep{Samus2017}, AKARI/HIP and AKARI/2MASS samples \citep{Ita2010}).

BY~CVn is also included in the 9$^{\rm th}$ Catalogue of Spectroscopic Binary Orbits \citep[SB9,][]{Pourbaix2004}. It is certainly a binary system with an orbital period of 496.7~days \citep{Famey2009}. A binary model of this object implies a1$\times$sini=6.62759e+07~km with a standard deviation (SD) of 2e+06~km and mass function f(m)=0.0471997~M$\odot$ with SD=0.0043~M$\odot$. These parameters are comparable with those obtained from known Systs \citep{fekel2000} or even binary systems with WD companion. We, hence, consider BY~CVn as a good SySt candidate for further investigation.

\begin{figure*}
\vbox{
\includegraphics[scale=0.375]{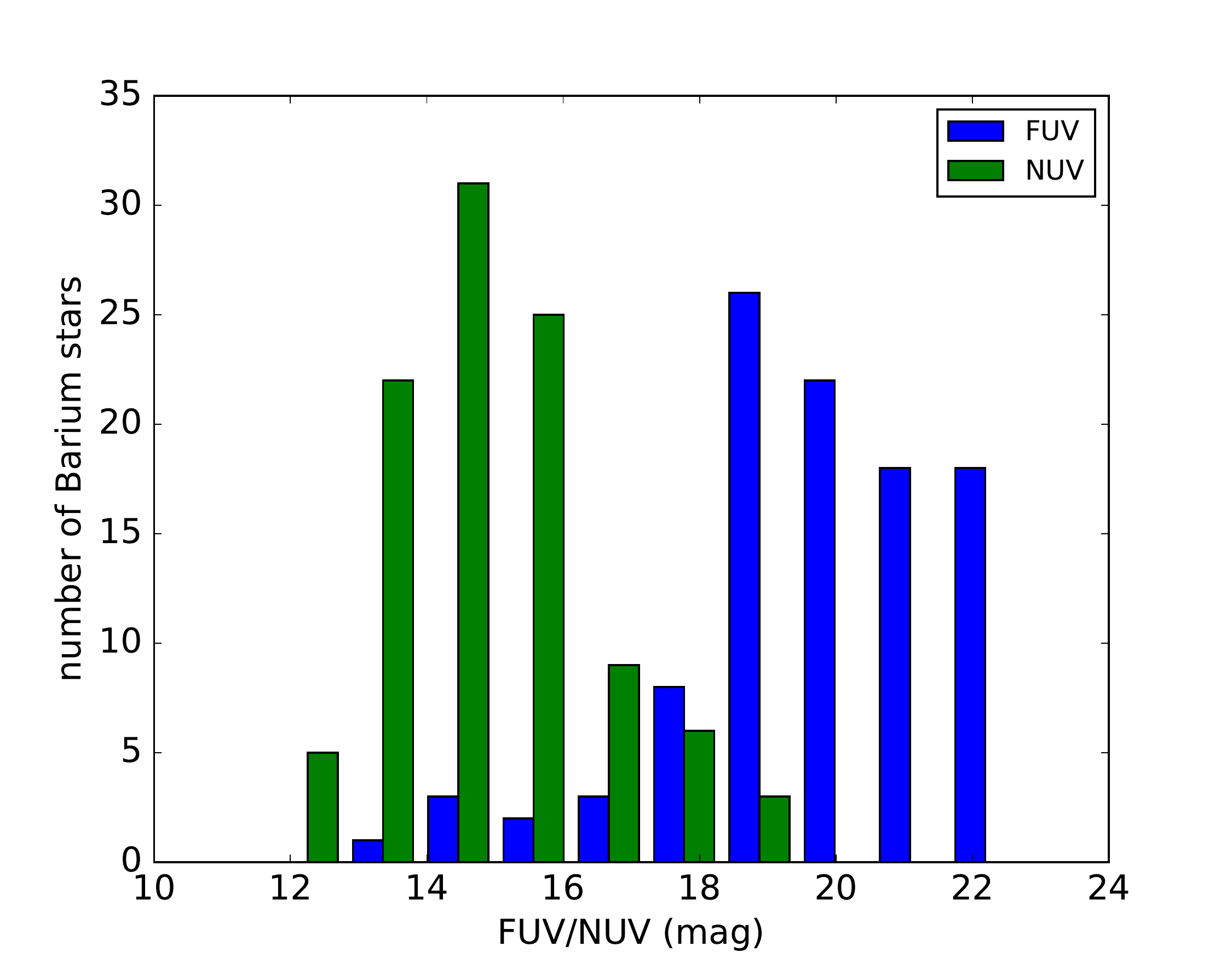}
\includegraphics[scale=0.375]{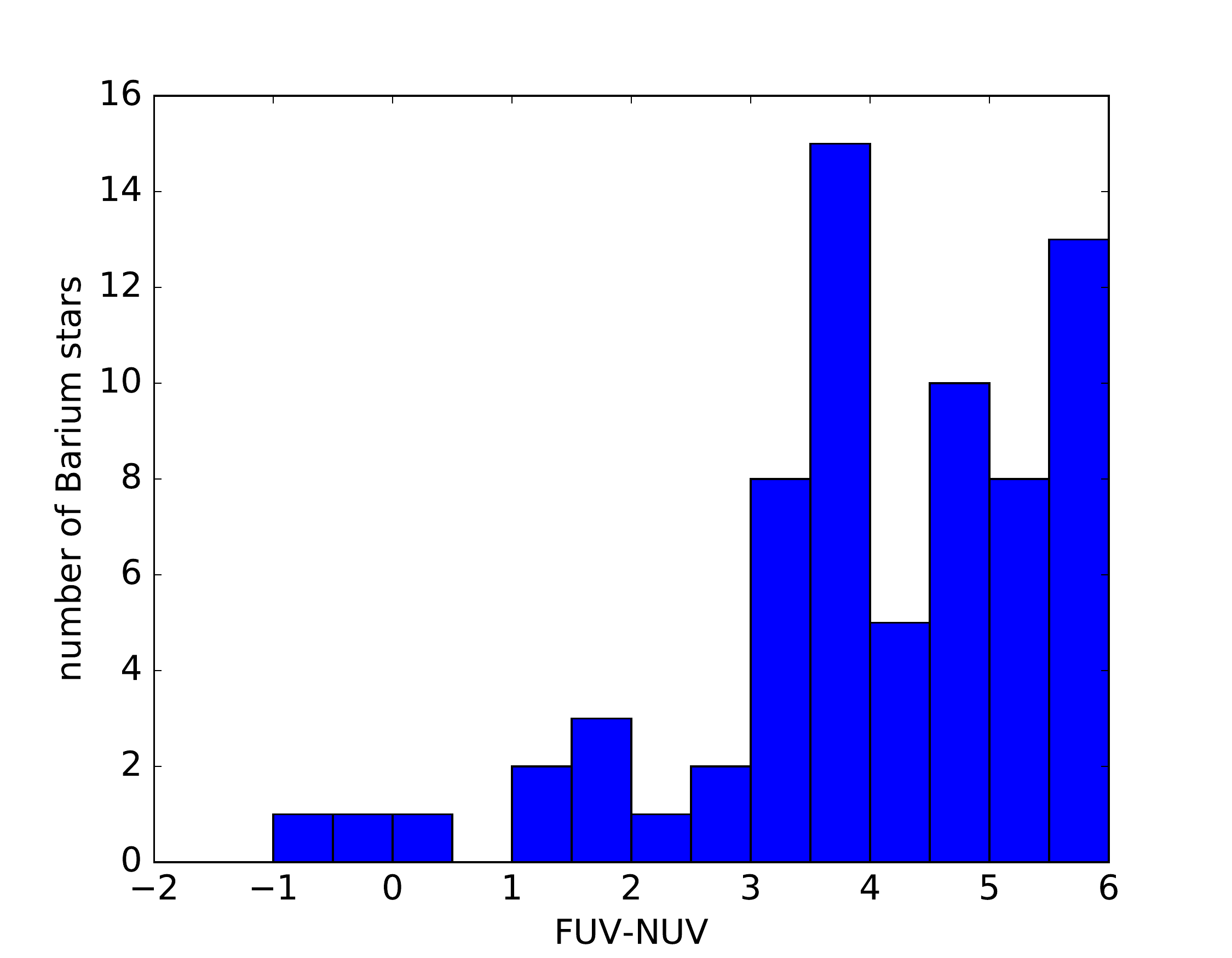}}
\caption[]{FUV/NUV photometry (left panel) and FUV-NUV colour index (right panel) distributions of Barium stars.}
\label{fig6}
\end{figure*}

\subsection{SySts versus Barium and Technetium-poor stars}

Besides SySts, our {\it GALEX/ASAS-SN} list may also contain other families of WD~+~RG binary systems such as Barium stars and Technetium-poor (Tc-poor) extrinsic S stars\footnote{For more information about these families of WD~+~RG binary systems, the reader should refer to \cite{McClure1980,Han1995, McClure1990,Jorissen2019})}. These two groups of stars exhibit peculiar abundances of s-process elements like Ba, Sr or Tc that cannot be explained from the evolutionary status of single stars. The process of mass transfer between the companions in binary systems has been proposed to explain these peculiarities \cite{McClure1980,Han1995, McClure1990,Jorissen2019}.

Barium stars have been found to be either dwarf or giant stars \citep{Escorza2017}. A UV-excess in Barium stars has been reported relative to normal giants \citep{BohmVitense2000,Gray2011} indicative of a hot compact companion. Most of them have been proved to be in binary systems \citep{McClure1980}, in which the more massive companion evolved faster, it produced the s-process elements in their interior during the Asymptotic Giant Branch (AGB) phase, brought them to the surface through the 2$^{nd}$ and 3$^{rd}$ dredges up processes, and finally contaminated the less massive and less evolved companion with the s-process elements through mass transfer. When the secondary is finally evolved to the giant phase, the primary has become a white dwarf.

The analysis of the IR and UV criteria in this work ensure the presence of a giant and WD companions, respectively. Therefore, Barium and extrinsic Tc-poor S stars may also be included in our {\it GALEX/ASAS-SN} list. To probe this hypothesis, we gathered a sample of Barium and extrinsic Tc-poor S stars from the literature in order to explore their distributions in IR and UV colours. Our lists contain 327 Barium and 50 Tc-poor S stars (a list of 30 Tc-rich intrinsic S stars is also used as a test sample).

Barium stars exhibit a mean 2MASS {\it J-H} colour index equal to 0.51 (SD=0.14) mag which is clearly smaller than the lower limit of SySts ({\it J-H}$>$0.78) and they can be easily distinguished. From the list of 327 Ba stars, only 14 (4 percent) satisfy the IR criteria of SySts \citep{Akras2019b}. Interestingly, all Barium stars violate the criterion that separates SySts from single red giants \citep[{\it K-W3}$>$0.27, see fig.~A5 in][]{Akras2019b}. Barium stars have a mean {\it K-W3} colour equal to 0.06 with SD=0.21. Regarding the Tc-poor S stars, 41 out of 50 Tc-poor S stars are found to satisfy all the IR criterion of SySts, but they also violate the criterion between SySts/RGs (({\it K-W3})$_{\rm mean}$=0.16, SD=0.22). The remaining nine Tc-poor S stars have {\it J-H}$<$0.78~mag. 

From the analysis above, it is clear that the possibility of finding a Barium or Tc-poor S stars in the {\it GALEX/ASAS-SN} list of candidates is nearly negligible. Both are well separated from SySts based on their infrared 2MASS/WISE colour indices, despite that they are all members of the WD~+~RG binary systems. SySts are characterized by higher {\it J-H}, {\it K-W3} and {\it W3-W4} colour indices compared to Barium and Tc-poor S stars, indicative of higher mass-loss rates. This might be one of the reasons that symbiotic activity is not triggered in most of the Barium and Tc-poor S stars.

{\it GALEX} UV photometry has also been gathered for our lists of Barium stars (101 have available measurements from both {\it GALEX} bands) and Tc-poor S stars (18). Figure~\ref{fig6} displays the distribution of the FUV and NUV photometry, as well as the FUV-NUV colour index for our sample of Barium stars. All but three Barium stars and all Tc-poor S stars display FUV-NUV$>$1. The mean of FUV-NUV colour index of Barium and Tc-poor S stars is calculated 4.8~mag (SD=1.7~mag) and 2.38~mag(SD=0.85~mag), respectively. Hence, both classes of WD~+~RG binary systems exhibit a distinct difference in the {\it GALEX} colour index from SySts. This is the second main difference, besides the IR colour indices, between SySts and Barium/Tc-poor S stars. 

It is worth mentioning that among the three Barium stars with FUV-NUV$<$1, there are two known SySts which also show the Barium phenomenon (BD-21 3873: FUV-NUV=-0.562~mag, LT~Del: FUV-NUV=-0.443~mag) and a variable star (29~Dra: FUV-NUV=0.312~mag). 29~Dra is a well-known WD+RG binary system \citep[K0~III, ][]{Fekel1993,Zboril2009,Bilikova2010,Pickles2010} with an orbital period of 903.8 days \citep{Fekel1993} and a rapidly rotating giant with strong chromospheric activity (strong Ca~II H and K emission lines). The giant star in this system is characterized by non-chemical abundances peculiarities \citep[e.g., ][]{Fekel1985,Zacs1997,Barisevicius2010,Merle2016} and positive FUV-NUV index. Moreover, it only marginally passes the criteria of S-type SySts (J-H=0.705$\pm$0.504, K-W3=-0.004$\pm$0.423, W1-W2=0.338$\pm$0.466, W1-W4=0.234$\pm$0.305) because of the large photometric error. The WD component has M=0.55~M$\odot$, slightly larger than the threshold of 0.5~M$\odot$~\citep{Merle2016}, T$_{\rm eff}$=30000~K and logg=8 \citep{Fekel1985}. It's FUV-NUV colour index of 0.312~mag is comparable with the colour index of SU~Lyn 0.323~mag. It is also characterized by hard X-ray emission \cite[e.g., ][]{Bilikova2010} like the $\delta$-type no shell-burning (accreting-only) SU~Lyn. Therefore, any potential symbiotic activity in this binary system may not be detectable through the typical optical emission lines criteria. 

The mean FUV-NUV colour index was also computed for the small sample of 9 Tc~rich stars \citep[][]{Jorrisen2016,Jorissen2019}, 22 carbon-enhanced, metal-poor stars (8 CEMP-no, 7 CEMP-rs, and 7 CEMP-s, \cite{Beers2005}), and 9 CH stars with available measurements in both {\it GALEX} filters. None of these types of stars is characterized by a low FUV-NUV colour index. Despite the small populations of our samples, their mean FUV-NUV colour indices are clearly higher than 1: Tc~rich: 3.02(SD=1.13)~mag, CEMP-no: 4.62(SD=1.07)~mag, CEMP-rs: 4.48(SD=1.71)~mag, CEMP-s: 4.41(SD=1.69)~mag, and CH: 6.17(SD=0.96)~mag.

All these classes of sources have been proposed to be members of binary systems with a WD companion \citep[][]{McClure1980,McClure1990,Han1995,Jorrisen2016,Jorissen2019}. Their high FUV-NUV colour indices, e.g. Barium and Tc-poor S stars, distinguish them from SySts. Their WD companions have significantly different stellar parameters than those in SySts or no accretion disk. For the connection between SySts and peculiar stars as well as the conditions necessary to provoke symbiotic activity and/or barium enhancement, the reader should refer to the review by \cite{Jorissen2002}.

\section{Conclusions}

In this work, I searched for new candidate SySts hidden in the ASAS-SN catalogue of variable stars~III, combining photometric data from GALEX, Gaia, 2MASS and AllWISE. The discovery of the accreting-only SU~Lyn and THA~15-31 SySts with weak or absent optical emission, intrigued me to search for SySts without take into account information from the visible.

{\it GALEX} UV photometry was useful to recover this hidden population of SySts unveiling the presence of the WDs in the spectral regime where red giants are not the dominant source. By applying the IR selection criteria of S-type SySts \citep{Akras2019b} to the ASAS-SN catalogue of variables stars, I obtained a list of potential SySts with RG companions (29,237) as it can be seen from their M$_{\rm G}$ magnitude and G$_{\rm BP}$-G$_{\rm RP}$ colour indices. The cross-match of that list with the revised catalogue of GALEX UV sources \citep{Bianchi2017} provided us with 814 matches, from which only 105 have available photometry from both {\it GALEX} bands. The remaining 709 sources have only NUV photometry and were considered only as potential candidate SySts.

Based on the FUV/NUV photometry as well as the FUV-NUV colour index of known SySts, it was found that the FUV-NUV$<$1 criterion is a reliable indicator for the presence of WDs. Moreover, the comparison of the {\it GALEX} photometry of WDs, M-dwarfs and M-giants demonstrated that the FUV-NUV$<$1 criterion yields mainly WDs with very low contamination from giant stars. The higher the FUV-NUV colour index, the higher the contamination from M-dwarfs and thus M-giant+M-dwarfs binary systems. 

The presence of two emission line stars (SS~305 and THA 23-30) in our list of {\it GALEX/ASAS-SN} candidates verified the method with follow-up optical spectroscopy. The symbiotic nature of both candidates is confirmed with the detection of the molecular bands and the strong \ha, \heliumb~$\lambda$4686, \ovi~$\lambda$6380 Raman-scattered lines, among others. 

These new discoveries strongly supported the combination of the IR, {\it GALEX} and Gaia information for searching potential SySt candidates. We also argued that many more SySts are likely hidden in our list of {\it GALEX/ASAS-SN} candidates (814) with only NUV photometry available. Considering that there are at least 7 genuine SySts in our list of 105 {\it GALEX/ASAS-SN} candidates with available FUV and NUV bands photometry, we reckoned that around 50 more SySts may be included in our list of {\it GALEX/ASAS-SN} candidates. Two more interesting sources with FUV-NUV$<$1 were found in the lists of M-type giants (BY CVn) and Barium stars (29 Dra).

\section*{Acknowledgements}
I would like to thank the anonymous Referee for the constructive comments that improved the paper. I would also like to thank Denise Rocha Gon\c{c}alves and Gerardo Juan Manuel Luna for their comments and suggestions on an earlier draft of this paper. This paper makes use of data from the ASAS-SN catalogue of Variable Stars~III, from the Two Micron All-Sky Survey which is a joint project of the University of Massachusetts and the Infrared Processing and Analysis Centre/California Institute of Technology, funded by the NASA and the National Science Foundation (NSF), and from the Wide-field Infrared Survey Explorer, which is a joint project of the University of California, Los Angeles, and the Jet Propulsion Laboratory/California Institute of Technology, funded by the NASA. This work has also made use of data from the European Space Agency (ESA) mission Gaia (\url{https://www.cosmos.esa.int/gaia}), processed by the Gaia Data Processing and Analysis Consortium (DPAC, \url{https://www.cosmos.esa.int/web/gaia/dpac/consortium}). Funding for the DPAC has been provided by national institutions, in particular the institutions participating in the Gaia Multilateral Agreement. SIMBAD database, operated at CDS, Strasbourg, France has also been used. Finally, the following software packages in Python were employed: Matplotlib \citep{Hunter2007}, NumPy \citep{Walt2011}, SciPy \citep{Virtanen2020} and AstroPy Python \citep{Astropy2013,Astropy2018}.

\section*{DATA AVAILABILITY}
The data and results of this article will be shared on reasonable request to the corresponding author.




\bibliographystyle{mnras}
\bibliography{references} 



\appendix

\bsp	
\label{lastpage}
\end{document}